\documentclass[10pt,a4paper,onecolumn]{article}
\usepackage{marginnote}
\usepackage{graphicx}
\usepackage{xcolor}
\usepackage{authblk,etoolbox}
\usepackage{titlesec}
\usepackage{calc}
\usepackage{tikz}
\usepackage{hyperref}
\hypersetup{colorlinks,breaklinks=true,
            urlcolor=[rgb]{0.0, 0.5, 1.0},
            linkcolor=[rgb]{0.0, 0.5, 1.0}}
\usepackage{caption}
\usepackage{tcolorbox}
\usepackage{amssymb,amsmath}
\usepackage{ifxetex,ifluatex}
\usepackage{seqsplit}
\usepackage{xstring}

\usepackage{float}
\let\origfigure\figure
\let\endorigfigure\endfigure
\renewenvironment{figure}[1][2] {
    \expandafter\origfigure\expandafter[H]
} {
    \endorigfigure
}

\usepackage{fixltx2e} 
\usepackage[
  backend=biber,
]{biblatex}
\bibliography{paper.bib}

\let\textttOrig=\texttt
\def\texttt#1{\expandafter\textttOrig{\seqsplit{#1}}}
\renewcommand{\seqinsert}{\ifmmode
  \allowbreak
  \else\penalty6000\hspace{0pt plus 0.02em}\fi}

\makeatletter
\let\href@Orig=\href
\def\href@Urllike#1#2{\href@Orig{#1}{\begingroup
    \def\Url@String{#2}\Url@FormatString
    \endgroup}}
\def\href@Notdoi#1#2{\def\tempa{#1}\def\tempb{#2}%
  \ifx\tempa\tempb\relax\href@Urllike{#1}{#2}\else
  \href@Orig{#1}{#2}\fi}
\def\href#1#2{%
  \IfBeginWith{#1}{https://doi.org}%
  {\href@Urllike{#1}{#2}}{\href@Notdoi{#1}{#2}}}
\makeatother

\newlength{\cslhangindent}
\newlength{\csllabelwidth}
\newenvironment{CSLReferences}[3] 
 {
  \setlength{\parindent}{0pt}
  \ifodd #1 \everypar{\setlength{\hangindent}{\cslhangindent}}\ignorespaces\fi
  \ifnum #2 > 0
  \setlength{\parskip}{#2\baselineskip}
  \fi
 }%
 {}
\usepackage{calc}

\usepackage[top=3.5cm, bottom=3cm, right=1.5cm, left=1.0cm,
            headheight=2.2cm, reversemp, includemp, marginparwidth=4.5cm]{geometry}

\titleformat{\section}
  {\normalfont\sffamily\Large\bfseries}
  {}{0pt}{}
\titleformat{\subsection}
  {\normalfont\sffamily\large\bfseries}
  {}{0pt}{}
\titleformat{\subsubsection}
  {\normalfont\sffamily\bfseries}
  {}{0pt}{}
\titleformat*{\paragraph}
  {\sffamily\normalsize}

\usepackage{fancyhdr}
\makeatletter
\let\ps@plain\ps@fancy
\fancyheadoffset[L]{4.5cm}
\fancyfootoffset[L]{4.5cm}

\definecolor{linky}{rgb}{0.0, 0.5, 1.0}

\newtcolorbox{repobox}
   {colback=red, colframe=red!75!black,
     boxrule=0.5pt, arc=2pt, left=6pt, right=6pt, top=3pt, bottom=3pt}

\newcommand{\ExternalLink}{%
   \tikz[x=1.2ex, y=1.2ex, baseline=-0.05ex]{%
       \begin{scope}[x=1ex, y=1ex]
           \clip (-0.1,-0.1)
               --++ (-0, 1.2)
               --++ (0.6, 0)
               --++ (0, -0.6)
               --++ (0.6, 0)
               --++ (0, -1);
           \path[draw,
               line width = 0.5,
               rounded corners=0.5]
               (0,0) rectangle (1,1);
       \end{scope}
       \path[draw, line width = 0.5] (0.5, 0.5)
           -- (1, 1);
       \path[draw, line width = 0.5] (0.6, 1)
           -- (1, 1) -- (1, 0.6);
       }
   }

\patchcmd{\@maketitle}{center}{flushleft}{}{}
\patchcmd{\@maketitle}{center}{flushleft}{}{}
\patchcmd{\@maketitle}{\LARGE}{\LARGE\sffamily}{}{}
\def\maketitle{{%
  
  \AB@maketitle}}
\makeatletter
\renewcommand\AB@affilsepx{ \protect\Affilfont}
\renewcommand\AB@affilnote[1]{{\bfseries #1}\hspace{3pt}}
\renewcommand{\affil}[2][]%
   {\newaffiltrue\let\AB@blk@and\AB@pand
      \if\relax#1\relax\def\AB@note{\AB@thenote}\else\def\AB@note{#1}%
        \setcounter{Maxaffil}{0}\fi
        \begingroup
        \let\href=\href@Orig
        \let\texttt=\textttOrig
        \let\protect\@unexpandable@protect
        \def\thanks{\protect\thanks}\def\footnote{\protect\footnote}%
        \@temptokena=\expandafter{\AB@authors}%
        {\def\\{\protect\\\protect\Affilfont}\xdef\AB@temp{#2}}%
         \xdef\AB@authors{\the\@temptokena\AB@las\AB@au@str
         \protect\\[\affilsep]\protect\Affilfont\AB@temp}%
         \gdef\AB@las{}\gdef\AB@au@str{}%
        {\def\\{, \ignorespaces}\xdef\AB@temp{#2}}%
        \@temptokena=\expandafter{\AB@affillist}%
        \xdef\AB@affillist{\the\@temptokena \AB@affilsep
          \AB@affilnote{\AB@note}\protect\Affilfont\AB@temp}%
      \endgroup
       \let\AB@affilsep\AB@affilsepx
}
\makeatother

\renewcommand\Affilfont{\sffamily\small\mdseries}
\ifnum 0\ifxetex 1\fi\ifluatex 1\fi=0 
  \usepackage[T1]{fontenc}
  \usepackage[utf8]{inputenc}

\else 
  \ifxetex
    \usepackage{mathspec}
    \usepackage{fontspec}

  \else
    \usepackage{fontspec}
  \fi
  \defaultfontfeatures{Ligatures=TeX,Scale=MatchLowercase}

\fi
\IfFileExists{upquote.sty}{\usepackage{upquote}}{}
\IfFileExists{microtype.sty}{%
\usepackage{microtype}
\UseMicrotypeSet[protrusion]{basicmath} 
}{}

\usepackage{hyperref}
\hypersetup{unicode=true,
            pdftitle={lattice-symmetries: A package for working with quantum many-body bases},
            pdfborder={0 0 0},
            breaklinks=true}
\let\addcontentslineOrig=\addcontentsline
\def\addcontentsline#1#2#3{\bgroup
  \let\texttt=\textttOrig\addcontentslineOrig{#1}{#2}{#3}\egroup}
\let\markbothOrig\markboth
\def\markboth#1#2{\bgroup
  \let\texttt=\textttOrig\markbothOrig{#1}{#2}\egroup}
\let\markrightOrig\markright
\def\markright#1{\bgroup
  \let\texttt=\textttOrig\markrightOrig{#1}\egroup}

\usepackage{graphicx,grffile}
\makeatletter
\def\maxwidth{\ifdim\Gin@nat@width>\linewidth\linewidth\else\Gin@nat@width\fi}
\def\maxheight{\ifdim\Gin@nat@height>\textheight\textheight\else\Gin@nat@height\fi}
\makeatother
\setkeys{Gin}{width=\maxwidth,height=\maxheight,keepaspectratio}
\IfFileExists{parskip.sty}{%
\usepackage{parskip}
}{
\setlength{\parindent}{0pt}
\setlength{\parskip}{6pt plus 2pt minus 1pt}
}
\ifx\paragraph\undefined\else
\let\oldparagraph\paragraph
\renewcommand{\paragraph}[1]{\oldparagraph{#1}\mbox{}}
\fi
\ifx\subparagraph\undefined\else
\let\oldsubparagraph\subparagraph
\renewcommand{\subparagraph}[1]{\oldsubparagraph{#1}\mbox{}}
\fi

\title{\texttt{lattice-symmetries}: A package for working with quantum
many-body bases}

        \author[1]{Tom Westerhout}
    
      \affil[1]{Institute for Molecules and Materials, Radboud
University}
  \date{\vspace{-7ex}}

\begin{document}
\maketitle

\marginpar{

  \begin{flushleft}
  \sffamily\small

  {\bfseries DOI:} \href{https://doi.org/DOI unavailable}{\color{linky}{DOI unavailable}}

  \vspace{2mm}

  {\bfseries Software}
  \begin{itemize}
    \setlength\itemsep{0em}
    \item \href{N/A}{\color{linky}{Review}} \ExternalLink
    \item \href{NO_REPOSITORY}{\color{linky}{Repository}} \ExternalLink
    \item \href{DOI unavailable}{\color{linky}{Archive}} \ExternalLink
  \end{itemize}

  \vspace{2mm}

  \par\noindent\hrulefill\par

  \vspace{2mm}

  {\bfseries Editor:} \href{https://example.com}{Pending
Editor} \ExternalLink \\
  \vspace{1mm}
    {\bfseries Reviewers:}
  \begin{itemize}
  \setlength\itemsep{0em}
    \item \href{https://github.com/Pending Reviewers}{@Pending
Reviewers}
    \end{itemize}
    \vspace{2mm}

  {\bfseries Submitted:} N/A\\
  {\bfseries Published:} N/A

  \vspace{2mm}
  {\bfseries License}\\
  Authors of papers retain copyright and release the work under a Creative Commons Attribution 4.0 International License (\href{http://creativecommons.org/licenses/by/4.0/}{\color{linky}{CC BY 4.0}}).

  \end{flushleft}
}

\hypertarget{summary}{%
\section{Summary}\label{summary}}

Exact diagonalization (ED) is one of the most reliable and established
numerical methods of quantum many-body theory. It is precise, unbiased,
and general enough to be applicable to a huge variety of problems in
condensed matter physics. Mathematically, ED is a linear algebra problem
involving a matrix called the Hamiltonian. For a system of spin-1/2
particles, the size of this matrix scales exponentially (as
\(\mathcal{O}(2^N)\)) with the number of particles \(N\).

Very fast scaling of memory requirements with system size is the main
computational challenge of the method. There are a few techniques
allowing one to lower the amount of storage used by the Hamiltonian. For
example, one can store only the non-zero elements of the Hamiltonian.
This is beneficial when the Hamiltonian is sparse, which is usually the
case in condensed matter physics. One can even take it one step further
and avoid storing the matrix altogether by instead computing matrix
elements on the fly.

A complementary approach to reduce memory requirements is to make use of
system symmetries. For example, many relevant Hamiltonians possess
\(U(1)\) symmetry, which permits one to perform calculations assuming
that the number of particles (or number of spins pointing upwards), is
fixed. Another example would be translational invariance of the
underlying lattice.

Although the algorithms for dealing with lattice symmetries are well
known (Sandvik et al., 2010), implementing them remains a non-trivial
task. Here we present \texttt{lattice-symmetries}, a package providing
high-quality and high-performance realization of these algorithms.
Instead of writing their own optimized implementation for every system
of interest, a domain expert provides system-specific details (such as
the number of particles or momentum quantum number) to
\texttt{lattice-symmetries} and it will automatically construct a
reduced Hamiltonian. The dimension of the new Hamiltonian can be
multiple orders of magnitude smaller than of the original one.

Furthermore, in \texttt{lattice-symmetries} the Hamiltonian itself is
never stored. Instead, its matrix elements are computed on the fly,
which reduces the memory requirements even more. Care is taken to keep
the implementation generic such that different physical systems can be
studied, but without sacrificing performance as we will show in the next
section.

All in all, \texttt{lattice-symmetries} serves as a foundation for
building state-of-the-art ED and VMC (Variational Monte Carlo)
applications. For example, \texttt{SpinED} (Westerhout, 2020) is an
easy-to-use application for exact diagonalization that is built on top
of \texttt{lattice-symmetries} and can handle clusters of at least 42
spins on a single node.

\hypertarget{statement-of-need}{%
\section{Statement of need}\label{statement-of-need}}

Exact diagonalization is an old and well-established method and many
packages have been written for it. However, we find that most
state-of-the-art implementations (Läuchli et al., 2019; Wietek \&
Läuchli, 2018) are closed-source. There are but three notable
open-source projects that natively support spin systems\footnote{There
  are a few projects targeting fermionic systems and the Hubbard model
  in particular. Although it is possible to transform a spin Hamiltonian
  into a fermionic one, it is impractical for large-scale simulations
  since lattice symmetries are lost in the new Hamiltonian.} :
\(\text{H}\Phi\) (Kawamura et al., 2017), \texttt{SPINPACK}
(Schulenburg, 2017), and \texttt{QuSpin} (Weinberg \& Bukov, 2017).

\begin{figure}
\centering
\includegraphics[width=0.9\textwidth,height=\textheight]{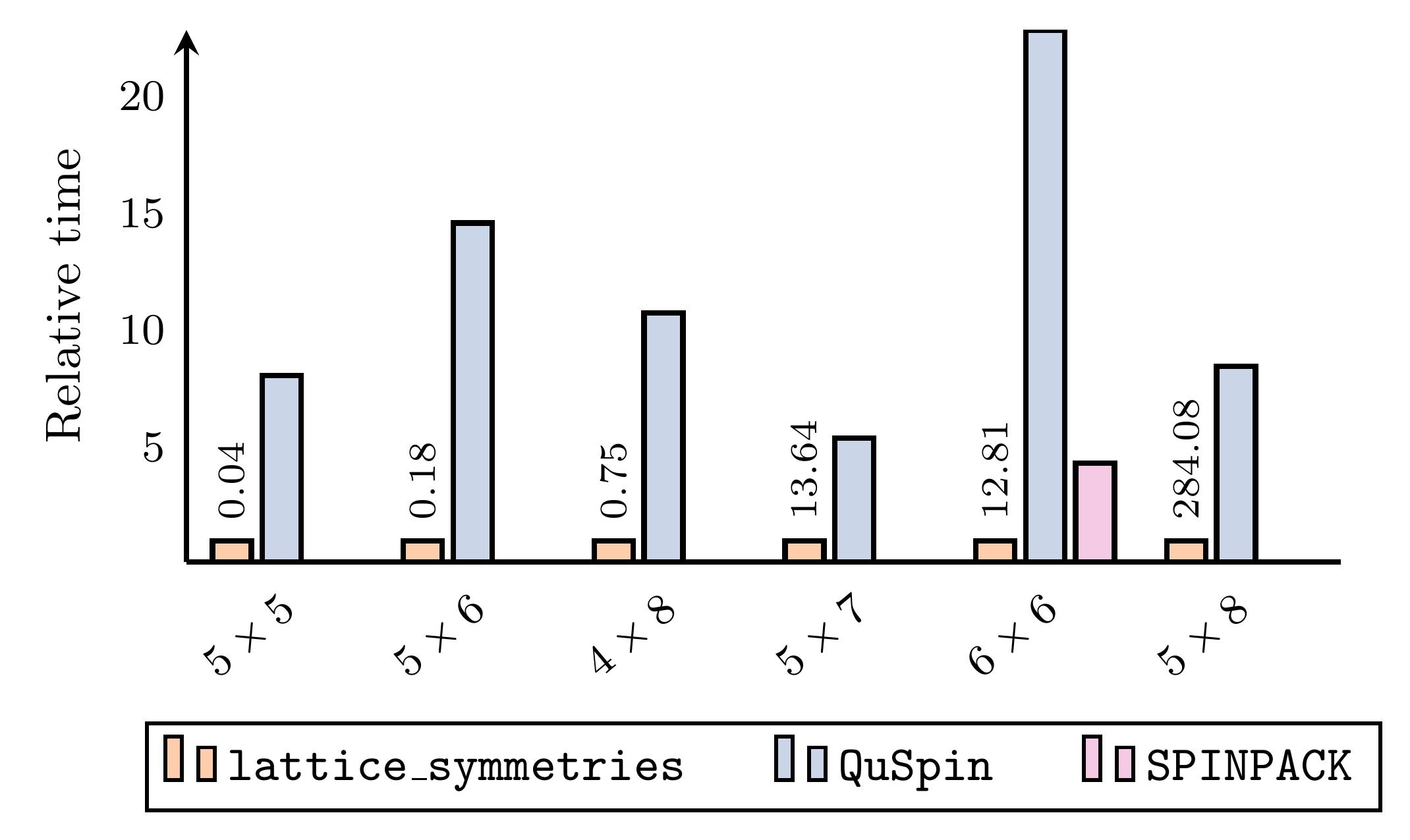}
\caption{Performance of matrix-vector products in \texttt{QuSpin},
\texttt{SPINPACK}, and \texttt{lattice-symmetries}. For a Heisenberg
Hamiltonian on square lattices of different sizes, we measure the time
it takes to do a single matrix-vector product. Timings for
\texttt{lattice-symmetries} are normalized to \(1\) to show relative
speedup compared to \texttt{QuSpin}, with absolute times in seconds
listed as well. Depending on the system speedups over \texttt{QuSpin}
vary between 5 and 22 times, but in all cases
\texttt{lattice-symmetries} is significantly faster.
\label{fig:performance}}
\end{figure}

\(\text{H}\Phi\) implements a variety of Hamiltonians, works at both
zero and finite temperatures, and supports multi-node computations.
However, there are a few points in which \texttt{lattice-symmetries}
improves upon \(\text{H}\Phi\). Firstly, \(\text{H}\Phi\) does not
support arbitrary lattice symmetries. Secondly, it uses a custom input
file format making it not user-friendly. Finally, since \(\text{H}\Phi\)
is an executable, it cannot be used to as a library to develop new
algorithms and applications.

\texttt{SPINPACK} is another popular solution for diagonalization of
spin Hamiltonians. \texttt{SPINPACK} does support user-defined
symmetries, unlike \(\text{H}\Phi\), but its interface is even less
user-friendly. Defining a lattice, Hamiltonian, and symmetries requires
writing non-trivial amounts of \texttt{C} code. Finally,
\texttt{SPINPACK} is slower than \texttt{lattice-symmetries} as
illustrated in \autoref{fig:performance}.

\texttt{QuSpin} is much closer in functionality to
\texttt{lattice-symmetries}. It is a high-level Python package, which
natively supports (but is not limited to) spin systems, can employ
user-defined lattice symmetries, and can also perform matrix-free
calculations (where matrix elements are computed on the fly). However,
\texttt{QuSpin} mostly focuses on ease of use and functionality rather
than performance. In \texttt{lattice-symmetries} we follow UNIX
philosophy (Salus, 1994) and try to ``do one thing but do it well.''
Even though \texttt{lattice-symmetries} uses essentially the same
algorithms as \texttt{QuSpin}, careful implementation allows us to
achieve an order of magnitude speedup as shown in
\autoref{fig:performance}. To achieve such performance, we make heavy
use of Single Instruction Multiple Data (SIMD) instructions supported by
modern processors. Vector Class Library (Fog, 2020) is used to write all
performance-critical kernels, which currently limits the portability of
\texttt{lattice-symmetries} to processors supporting \texttt{x86-64}
instruction sets.

\texttt{lattice-symmetries} is a library implemented in \texttt{C++} and
\texttt{C}. It provides two interfaces:

\begin{itemize}
\item
  Low-level \texttt{C} interface which can be used to implement ED and
  VMC applications with focus on performance.
\item
  A higher-level \texttt{Python} wrapper which allows to easily test and
  prototype algorithms.
\end{itemize}

The library is easily installable via the \texttt{Conda} package
manager.

The general workflow is as follows: the user starts by defining a few
symmetry generators (\texttt{ls\_symmetry}/\texttt{Symmetry} in
\texttt{C}/\texttt{Python}) from which \texttt{lattice-symmetries}
automatically constructs the symmetry group
(\texttt{ls\_group}/\texttt{Group} in \texttt{C}/\texttt{Python}). The
user then proceeds to constructing the Hilbert space basis
(\texttt{ls\_spin\_basis}/\texttt{SpinBasis} in
\texttt{C}/\texttt{Python}). For some applications functionality
provided by \texttt{SpinBasis} may be sufficient, but typically the user
will construct one (or multiple) quantum mechanical operators
(\texttt{ls\_operator}/\texttt{Operator} in \texttt{C}/\texttt{Python})
corresponding to the Hamiltonian and various observables.
\texttt{lattice-symmetries} supports generic 1-, 2-, 3-, and 4-point
operators. Examples of Hamiltonians that can be constructed include

\[
\begin{aligned}
    H &= \sum_{i, j} J_{ij} \boldsymbol\sigma_i \cdot \boldsymbol\sigma_j && \text{Heisenberg interaction,} \\
    H &= \sum_{i, j} J_{ij} \sigma^z_i \sigma^z_j + \sum_i h_i \sigma^x_i && \text{Ising model in transverse magnetic field,} \\
    H &= \sum_{i, j} \mathbf{D}_{ij} \left[ \boldsymbol\sigma_i \times \boldsymbol\sigma_j \right] && \text{Dzyaloshinskii-Moriya interaction.} \\
\end{aligned}
\]

Here, \(\boldsymbol\sigma\) denotes Pauli matrices, \(J\) and
\(\mathbf{D}\) are various coupling constants, and sums over \(i\) and
\(j\) can run over arbitrary used-defined geometries.

\texttt{Operator}s can be efficiently applied to vectors in the Hilbert
space (i.e., wavefunctions). Also, in cases when the Hilbert space
dimension is so big that the wavefunction cannot be written down
explicitly (as a list of coefficients), \texttt{Operator} can be applied
to individual spin configurations to implement Monte Carlo local
estimators.

As an example of what can be done with \texttt{lattice-symmetries}, we
implemented a standalone application for exact diagonalization studies
of spin-1/2 systems, \texttt{SpinED}. By combining
\texttt{lattice-symmetries} with the PRIMME eigensolver (Stathopoulos \&
McCombs, 2010), it allows one to treat systems of at least 42 sites on a
single node. \texttt{SpinED} is distributed as a statically-linked
executable --- one can download one file and immediately get started
with physics. All in all, it makes large-scale ED more approachable for
non-experts.

Notable research projects using \texttt{lattice-symmetries} and
\texttt{SpinED} include Astrakhantsev et al. (2021), Bagrov et al.
(2020), and Westerhout et al. (2020). Additionally,
\texttt{lattice-symmetries} is being used to simulate quantum circuits
(Astrakhantsev, 2021).

\hypertarget{acknowledgements}{%
\section{Acknowledgements}\label{acknowledgements}}

The author thanks Nikita Astrakhantsev for useful discussions and
testing of (most) new features. The assistance of Askar Iliasov and
Andrey Bagrov with with group-theoretic aspects of this work is
appreciated. This work was supported by European Research Council via
Synergy Grant 854843 -- FASTCORR.

\hypertarget{references}{%
\section*{References}\label{references}}
\addcontentsline{toc}{section}{References}

\hypertarget{refs}{}
\begin{CSLReferences}{1}{0}
\leavevmode\hypertarget{ref-qsl2021}{}%
Astrakhantsev, N. (2021). {QSL\_at\_QC}. In \emph{GitHub repository}.
\url{https://github.com/nikita-astronaut/QSL_at_QC}; GitHub.

\leavevmode\hypertarget{ref-astrakhantsev2021}{}%
Astrakhantsev, N., Westerhout, T., Tiwari, A., Choo, K., Chen, A.,
Fischer, M. H., Carleo, G., \& Neupert, T. (2021). Broken-symmetry
ground states of the {Heisenberg} model on the pyrochlore lattice.
\emph{arXiv Condensed Matter}. \url{http://arxiv.org/abs/2101.08787}

\leavevmode\hypertarget{ref-bagrov2020}{}%
Bagrov, A. A., Iliasov, A. A., \& Westerhout, T. (2020). Kinetic
samplers for neural quantum states. \emph{arXiv Condensed Matter}.
\url{http://arxiv.org/abs/2011.02986}

\leavevmode\hypertarget{ref-vectorclass}{}%
Fog, A. (2020). {VCL:} {C++} vector class library. In \emph{GitHub
repository}. \url{https://github.com/vectorclass/version2}; GitHub.

\leavevmode\hypertarget{ref-kawamura2017}{}%
Kawamura, M., Yoshimi, K., Misawa, T., Yamaji, Y., Todo, S., \&
Kawashima, N. (2017). Quantum lattice model solver {H\(\Phi\)}.
\emph{Computer Physics Communications}, \emph{217}, 180--192.
\url{https://doi.org/10.1016/j.cpc.2017.04.006}

\leavevmode\hypertarget{ref-lauchi2019}{}%
Läuchli, A. M., Sudan, J., \& Moessner, R. (2019). \(S=\frac{1}{2}\)
kagome {Heisenberg} antiferromagnet revisited. \emph{Physical Review B},
\emph{100}(15, 15), 155142.
\url{https://doi.org/10.1103/physrevb.100.155142}

\leavevmode\hypertarget{ref-salus1994}{}%
Salus, P. H. (1994). \emph{A quarter century of {UNIX}}. ACM
Press/Addison-Wesley Publishing Co.

\leavevmode\hypertarget{ref-sandvik2010}{}%
Sandvik, A. W., Avella, A., \& Mancini, F. (2010). Computational studies
of quantum spin systems. \emph{AIP Conference Proceedings}, \emph{1297},
135--338. \url{https://doi.org/10.1063/1.3518900}

\leavevmode\hypertarget{ref-schulenburg2017}{}%
Schulenburg, J. (2017). {SPINPACK}. \emph{Magdeburg University}.
\url{https://www-e.ovgu.de/jschulen/spin/}

\leavevmode\hypertarget{ref-stathopoulos2010}{}%
Stathopoulos, A., \& McCombs, J. R. (2010). {PRIMME}: Preconditioned
iterative multimethod eigensolver --- methods and software description.
\emph{ACM Transactions on Mathematical Software}, \emph{37}(2), 1--30.
\url{https://doi.org/10.1145/1731022.1731031}

\leavevmode\hypertarget{ref-weinberg2017}{}%
Weinberg, P., \& Bukov, M. (2017). {QuSpin:} {A} {Python} package for
dynamics and exact diagonalisation of quantum many body systems part
{I}: {Spin} chains. \emph{SciPost Physics}, \emph{2}(1, 1), 003.
\url{https://doi.org/10.21468/scipostphys.2.1.003}

\leavevmode\hypertarget{ref-SpinED}{}%
Westerhout, T. (2020). {SpinED}. In \emph{GitHub repository}.
\url{https://github.com/twesterhout/spin-ed}; GitHub.

\leavevmode\hypertarget{ref-westerhout2020}{}%
Westerhout, T., Astrakhantsev, N., Tikhonov, K. S., Katsnelson, M. I.,
\& Bagrov, A. A. (2020). Generalization properties of neural network
approximations to frustrated magnet ground states. \emph{Nature
Communications}, \emph{11}(1), 1--8.
\url{https://doi.org/10.1038/s41467-020-15402-w}

\leavevmode\hypertarget{ref-wietek2018}{}%
Wietek, A., \& Läuchli, A. M. (2018). Sublattice coding algorithm and
distributed memory parallelization for large-scale exact
diagonalizations of quantum many-body systems. \emph{Physical Review E},
\emph{98}(3, 3), 033309.
\url{https://doi.org/10.1103/physreve.98.033309}

\end{CSLReferences}

\end{document}